\documentclass{ws-p8-50x6-00}

\begin{document}

\title{The Formation of the Milky Way in the Cosmological Context}

\author{A. Burkert}

\address{Max-Planck-Institut f\"ur Astronomie, K\"onigstuhl 17, D-69117 Heidelberg,
Germany\\E-mail: burkert@mpia-hd.mpg.de}


\maketitle

\abstracts{
The formation of the Milky Way is discussed within the context of the cold dark matter 
scenario.
Several problems arise which can be solved if the Galaxy experienced an early phase of gas
heating and decoupling from the dark matter substructure. This model combines the
Eggen, Lynden-Bell and Sandage picture of a monolithic protogalactic collapse
with the Searle and Zinn picture of an early merging phase of substructures
into one consistent scenario of Galactic formation.}

\section{ELS versus Searle \& Zinn}

Because of its proximity, the Milky Way Galaxy is an ideal candidate 
to investigate the formation and evolution of galaxies. In 1962 one of the most influential
papers on galaxy formation was published by Eggen, Lynden-Bell and Sandage who
studied the motions of a sample of high-velocity stars. They found a correlation between metallicity and
orbital eccentricities and angular momentum and inferred that the Galaxy formed quickly, through the collapse 
of a uniform, isolated protogalactic cloud. 
In 1978, this scenario was challenged by Searle and Zinn who analysed Galactic globular clusters and
concluded from the lack of a metallicity gradient and from a large age spread that the halo was built 
on a timescale of several Gyrs from independent low-mass fragments.

Subsequent surveys, coupled with more precise determinations of
stellar ages and stellar element abundances have revealed in greater details the chemodynamical formation 
history of the Galaxy. No significant metallicity gradient has been found in the Galactic halo,
confirming the predictions by Searle and Zinn. 
In addition, strong evidence for a merging history of the Galaxy came from coherent moving groups of halo stars,
that might represent remnants of the original substructures (Majewski et al. 1994, Helmi et al. 1999).

Element abundances have also provided constraints on the timescale of early Galactic 
nucleosynthesis and by this on the evolution of the Milky Way
(Wheeler, Sneden \& Truran 1989). In the Galactic halo, oxygen and intermediate mass nuclei are enriched
by a factor of order 3, relative to iron. This is characteristic of pure high-mass star contamination by
supernovae Type II with lifetimes of less than $10^8$ yrs (Tinsley 79, Truran 91). 
Deviations from this trend seem first to
appear at the point where the first disk stars appear, indicating the onset of substantial enrichment
from Type Ia supernovae (Matteucci \& Greggio 86). Unfortunately, the lifetime of Type Ia supernova
progenitors is still poorly understood and might even depend on metallicity (Kobayashi et al. 98).
These objects therefore cannot be used as clocks to measure the timescale of disk formation.
One interesting coincidence is however the similarity of the epoch where SN Ia  begin to  contribute
to the metal enrichment in the Milky Way and the formation epoch of the galactic disk.
It is not clear why these two very different processes should occur on the same timescale.

Truran (1981) was the first to suggest that  low $[Ba/Eu]$-ratios of very
metal-deficient halo stars with $[Fe/H] < -2.5$ reflect 
a pure r-process contamination.  At a
metallicity of $[Fe/H] \approx -2.5$, the ratio of s-process/r-process, as reflected by 
[Ba/Eu] increases which indicates the entry of products from AGB stars, corresponding to
lifetimes of order $10^9$ yrs. If one assumes a typical disk lifetime of order 10-11 Gyrs, as inferred
from the white dwarf luminosity function and a galactic age  of 13 Gyrs as predicted by cosmological models,
the galactic halo  formed on timescales of $2 \times 10^9$ yrs with a roughly 
constant SN II rate of order 0.01 $yr^{-1}$ which is consistent with the Searle \& Zinn model.

\section{Cosmological predictions}

Up to now, the formation of the Milky Way has been studied neglecting its cosmological formation history.
Recently however, cosmological simulations have been able to resolve structure formation on galactic 
scales. For the first time we are now able to construct
a self-consistent model of the formation of the Milky Way, starting from realistic,
cosmologically motivated initial conditions.  The standard cold dark matter (CDM) scenario 
predicts that the Galaxy formed as suggested by Searle \& Zinn from many small substructures on
timescales of order $10^9$ yrs, in agreement
with the observational evidence. 

One of the fundamental assumptions of cold dark matter is its collisionless nature and the fact that
the CDM particles formed with negligible random velocities and subsequently interacted only 
through gravitational forces.
Numerous simulations have by now shown that this assumption leads to dark matter halos which have
universal density distributions that can  be approximated well by the simple function
(Navarro, Frenk \& White 1997; Moore et al. 1999)
$\rho(r) = \rho_0/ \left(r^{1.5}(r+r_s)^{1.5} \right)$ where $r_s$ is the scale radius which for the Galaxy is
of order 10 kpc. The Galactic disk therefore lies within the central
power-law cusp of the Milky Way's dark halo.
This universal density structure is not sensitive to the adopted cosmological model and is a universal property
of CDM.

Interestingly, this strong prediction of CDM models leads to several unsolved problems. 
(i) Observations of dark matter dominated
dwarf galaxies indicate that their dark halos have isothermal, constant density cores instead of
compact power-law cusps (Moore 1994, Burkert 1995). 
(ii) Due to their compact cores, the substructures from which halos form are not completely
disrupted, leading to the cosmological prediction that galaxies like the Milky Way should have of order 
a thousand satellites (Klypin et al. 99, Moore et al. 99). Only a dozen satellites
have been found so far. (iii) Cosmological simulations, including gas dynamics
predict a disk scale length for the Milky Way of order 300 pc, whereas its observed scale length is a factor
of 10 larger (Steinmetz \& Navarro 1999). 
Again, this problem is related to dense substructures from which 
the Galaxy formed. As their cores are not disrupted during the merging epoch they experience dynamical
friction by this loosing 90\% of their angular momentum. The gas cools and is
trapped in these cores. Dynamical friction will therefore also remove 
90\% of the angular momentum of the gaseous component, 
leading eventually to galactic disks with very small scale lenghts.

In summary, although cosmological models can by now resolve structure formation on
galactic scales, the dark matter core problem makes
it difficult to construct a consistent model of the formation of the Galaxy.

\section{Evidence for an Early Heating Phase of the Milky Way}

One solution of the cosmological satellite and disk problem  is an early phase of gas heating and decoupling
from the dark matter substructure (Weil et al. 1998; Sommer-Larsen et al. 1999). 
Suppose that gas was settling into the low-mass  progenitors of
the Milky Way. A first stellar population II formed at their center. Subsequent energetic feedback processes ejected 
the gas in a violent galactic wind. If the gas could be kept from falling back, the gas-free, dark  substructures
could merge, forming the dark halo of the Milky Way. The population II stars in their cores would now represent
the halo component. As star formation in the substructures was very inefficient, most
of the satellites would not easily be detectable. Their stars could represent some of  the moving groups
that have been found by recent surveys. The hot gas component, on the other hand, 
cooled and settled into the equatorial plane through a
monolithic collapse phase, conserving its angular momentum and forming a disk-like substructure
with the observed scale length. During the first phase of disk formation
the dark halo was still violently relaxing with massive substructures frequently interacting with
the proto-disk. During this early phase which might have lasted a Gyr the thick disk
formed. Subsequently, after the major merging epoch, gas could settle into a thin disk. 

This scenario can in fact explain an interesting puzzle that has been discussed by Wyse \& Gilmore (92). 
They noted that the specific angular momentum of the Galactic halo stars is a factor of 10 smaller than
required to form the Galactic disks. As a result, the halo could not have been the progenitor of the
disk. Its metal-enriched gas should instead have settled into the Galactic bulge. 
This seems to be in contradiction with the bulge mass and with the chemical continuity between the halo stars and 
the thick and thin disk component. If on the other hand 
the gas was ejected by a first population of stars before the major merging
event, the gas could keep its initial angular momentum whereas the halo stars, residing inside the cores
of the dark matter substructures would experience angular momentum loss by dynamical friction as predicted
by the cosmological models, leading to a low-angular momentum halo and a disk with large angular momentum.

The ELS and Searle \& Zinn picture have been considered as contradicting theories in the past.
An early phase of galactic heating would however combine both models in a new, consistent model
of Galaxy formation.  Whereas the galactic halo formed by
hierarchical merging of substructures, the Galactic thick and thin disk components resulted from
the monolithic infall of a diffuse gaseous component that was ejected from the substructures prior
to the merging event.

\section*{Acknowledgments}
I would like to thank the LOC for the excellent organization of an exciting conference.

\end{document}